\documentclass[twocolumn,showpacs,prl]{revtex4-1}
\usepackage{amssymb}
\usepackage{graphics}
\usepackage{epsfig}
\usepackage{dcolumn}
\usepackage{amsmath}
\usepackage{color}

\newcommand{\beq}{\begin{equation}}
\newcommand{\eeq}{\end{equation}}
\newcommand{\bea}{\begin{eqnarray}}
\newcommand{\eea}{\end{eqnarray}}
\newcommand{\bwt}{\begin{widetext}}
\newcommand{\ewt}{\end{widetext}}

\begin{document}

\title{Effect of Electron-electron Interaction on Surface Transport in Three-Dimensional Topological Insulators}
\date{\today}
\author{H. K. Pal,$^1$, V. I. Yudson$^2$, and D. L. Maslov$^1$}

\begin{abstract}
We study the effect of electron-electron interaction on 
the
surface 
resistivity
of three-dimensional 
(3D)
topological insulators. 
In the absence of umklapp scattering, the 
 existence 
  of the Fermi-liquid ($T^2$) term 
 in resistivity of a two-dimensional (2D) metal 
  depends on the Fermi surface geometry, in particular, on whether it is convex or concave.
 On doping, the Fermi surface of 2D metallic 
surface
states in 
3D
topological insulators of the Bi$_2$Te$_3$ family 
  changes its shape from convex to concave 
  due to hexagonal warping, while still being too small to allow for umklapp scattering.  
 We show that the $T^2$ term in the resistivity is present only in the concave regime and 
demonstrate
that the resistivity obeys a universal scaling form valid for an arbitrary 2D Fermi surface near a convex/concave transition.

\end{abstract}


\affiliation{
$^1$Department of Physics, University of Florida, Gainesville, FL 32611-8440, USA\\
$^2$Institute for Spectroscopy, Russian Academy of Sciences, Troitsk, Moscow Region, 142190, Russia}

\maketitle

Topological insulators 
(TI)
are characterized by a gapped bulk spectrum with conducting surface states extending across the entire gap. The surface states contain an odd number of Dirac cones and are protected against any perturbation that preserves time-reversal symmetry \cite{general}.
A wide variety of interesting physics resulting from these surface states is expected to be observed ranging from Majorana fermions \cite{majorana} to magnetic monopoles \cite{monopole}. Although 
photoemission and tunneling microscopy \cite{arpes_stm} have convincingly established the presence of such surface states in these materials, signatures of these states in transport measurements are more difficult to observe, mainly because of 
strong conduction in the bulk
 \cite{bulk}.
 With recent experimental progress, however, in the ability to tune the number of surface charge carriers \cite{tune}, it is now possible to see more clearly evidence of surface transport. In light of
this progress, it is timely to ask what is the effect of 
the electron-electron ({\em e-e}) 
interaction 
on surface transport.
Indeed, including {\em e-e} interaction 
is crucial for explaining the observed field and temperature dependences in quantum magnetotransport 
\cite{magres}. In this Letter, we address the manifestation of the {\em e-e} interaction in semiclassical
 transport within a model of a two-dimensional Fermi liquid relevant for surface states doped away from the Dirac point.

An archetypal signature of the Fermi-liquid behavior in metals
is the $T^2$ dependence of the resistivity ($\rho$).
With an exception of compensated semi-metals \cite{baber:37}, this dependence in clean conductors arises due to a special type of scattering processes--``umklapps" \cite{peierls:29,landau:36}--
in which the total momentum of an electron pair is changed
 by an integer multiple of the reciprocal lattice vector. Umklapps are possible
if certain conditions are met, namely, if the Fermi surface (FS) is large enough (the band is more than quarter full) and if the 
matrix element of the interaction has sufficient weight at large momentum transfers. Otherwise, the umklapp
contribution to the resistivity is suppressed.
 In this case, the $T^2$ contribution to $\rho$ may still occur due to the combined effect of the momentum-conserving (``normal") interaction among electrons on a lattice  and electron-impurity ({\em e-i})  scattering. 
Whether this really happens, turns  out to depend crucially on the dimensionality.  While the $T^2$ term is allowed for an 
anisotropic
FS 
with a non-parabolic spectrum
in 
three dimensions
(3D), the conditions in 
two dimensions
(2D)
are much more stringent \cite{gurzhi}. In particular, a $T^2$ term occurs in 2D only if the FS is either concave or multiply-connected; otherwise,   the leading {\em e-e} contribution scales as $T^4$ \cite{maslov}. Likewise, the frequency dependence of the 
{\em ac}
resistivity scales as $\Omega^4$ instead of $\Omega^2$ \cite{rosch}.
The reason for such a behavior is that the $T^2$ ($\Omega^2$) term arises from electrons confined to 
the FS contour. For a convex and singly-connected contour, the momentum and energy conservations 
are
similar to the 1D case,
where no relaxation is possible.

\begin{figure}
\includegraphics[width=30mm]{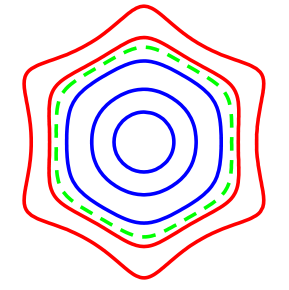}
\caption{(color online). Isoenergetic contours for the spectrum in Eq.~(\ref{eq:spectrum}). The dashed  line corresponds to the critical energy for the convex/concave transition.}
\label{fig1}
\end{figure}

We propose the surface state of a 3D TI as a testing ground for the theoretical results outlined above. Photoemission 
shows that the surface states of the Se, Te based compounds (Bi$_2$Te$_3$, Bi$_2$Se$_3$, and Sb$_2$Te$_3$) have a small, singly-connected FS at the center of the Brilllouin zone (BZ). 
Following Fu \cite{fu}, the electronic dispersion in these systems can be described by 
\beq
\epsilon_{\mathbf{k}}^{\pm}=\pm\sqrt{v^2k^2+\lambda^2 k^6\mathrm{cos}^2(3\theta)},
\label{eq:spectrum}
\eeq
where $\theta$ is the azimuthal angle, $v$ is the Dirac velocity, and $\lambda$ is a constant.
Corresponding isoenergetic  
 contours are presented in Fig.~\ref{fig1}. 
As the Fermi energy increases, the FS changes rapidly from a circle to a hexagon and then to a hexagram. At some critical value of the Fermi energy $\epsilon_F=\epsilon_c$ ($=0.16$\;eV for Bi$_2$Te$_3$, for example \cite{fu}),  the shape changes from convex to concave. Theory \cite{rosch,maslov} predicts, therefore, that the {\em e-e} contribution to the resistivity scales as $\max\{T^4,\Omega^4\}$ on the convex side and as $\max\{T^2,\Omega^2\}$ on the concave side.
The main result of this Letter is that,  near the convex/concave transition, the resistivity obeys a universal scaling form
\beq
\rho=\rho_0 + A\left(\frac{\Delta}{\epsilon_F}\right)^{9/2}\Theta (\Delta) T^2 + BT^4,
\label{eq:result}
\eeq
where  $\rho_0$ is the residual resistivity, $\Delta=\epsilon_F-\epsilon_c$, $\theta(x)$ is the step function, and $A$, $B$ are material-dependent parameters. 

The  exponents of  $2$, $4$, and $9/2$ in Eq.~(\ref{eq:result}) are universal, i.e., they are the same for 
an {\em arbitrary} 2D Fermi surface with a non-quadratic energy spectrum near a convex/concave transition. We emphasize, however, that the surface states of 3D TIs present a unique case of a small yet strongly warped 2D FS, where the predicted effects can be seen best. The drawback of 3D TIs is that they
have a 
large background dielectric constant ($\epsilon\sim 29-85$\;\cite{dielectric}), 
and hence, the electron-phonon ({\em e-ph}) 
interaction
is expected to dominate the $T$-dependence of the resistivity down to very low $T$\cite{eph}. This drawback can be circumvented by measuring the frequency dependence of the optical conductivity at frequencies above the Bloch-Gruneisen frequency, where the {\em e-e} contribution dominates over the  {\em e-ph} one \cite{ephsaturate}.

As in Ref.~\cite{maslov}, we adopt an approach based on the semiclassical Boltzmann equation (BE) \cite{comment2} and neglect quantum corrections to the conductivity.  For simplicity, the {\em e-i} interaction is accounted for within the
$1/\tau$ approximation. 
First, we consider the {\em dc} case ($\Omega=0$). 
For low enough $T$, when $\tau_{ee}\gg\tau_{ei}$, 
we solve the BE to leading order in the {\em e-e} interaction and obtain 
the correction to the residual conductivity as
\bea
\delta\sigma_{jj}=&-&\frac{e^2\tau_{i}^2}{2T}\int\frac{d^2q}{(2\pi)^2}\int\int\int d\omega d\epsilon_{\mathbf{k}}d\epsilon_{\mathbf{p}}\oint\oint\frac{da_{\mathbf{k}}}{v_{\mathbf{k}}}\frac{da_{\mathbf{p}}}{v_{\mathbf{p}}}\nonumber \\
&\times&|M_{\mathbf{k},\mathbf{p}}(\mathbf{q}
)|^2(\Delta \mathbf{v}_j)^2n(\epsilon_{\mathbf{k}})n(\epsilon_{\mathbf{p}})[1-n(\epsilon_{\mathbf{k}}-\omega)]\label{delsigma}\\
&\times&[1-n(\epsilon_{\mathbf{p}}+\omega)]\delta(\epsilon_{\mathbf{k}}-\epsilon_{\mathbf{k}-\mathbf{q}}-\omega)
\delta(\epsilon_{\mathbf{p}}-\epsilon_{\mathbf{p}+\mathbf{q}}+\omega).
\nonumber
\eea
Here, $\tau_{i}$ 
is the mean free time due 
to impurity scattering,
$\mathbf{q}$ and $\omega$ are the momentum and energy transfers,  
$\Delta\mathbf{v}=\mathbf{v}_{\mathbf{k}}+\mathbf{v}_{\mathbf{p}}-\mathbf{v}_{\mathbf{k}-\mathbf{q}}-\mathbf{v}_{\mathbf{p}+\mathbf{q}}$, $da_{\mathbf{l}}$ is the FS element, 
$n(\epsilon_{\bf l})=\left(\exp\left(\epsilon_{\bf l}/T\right)+1\right)^{-1}$, and $M_{\mathbf{k},\mathbf{p}}(\mathbf{q})$ is the matrix element of the {\em  e-e} interaction. 
To obtain the lowest in $T$ term in $\delta\sigma_{jj}$,  we project electrons onto the FS, which amounts to neglecting $\omega$ in the arguments of the $\delta$ functions. Since the typical values of $\epsilon_{\bf k,p}$ and $\omega$ are of order $T$, while the typical values of other variables are $T$ independent, $\delta\sigma_{jj}$ scales as $T^2$, which is the expected Fermi-liquid behavior. However, whether the prefactor of the $T^2$ term is non-zero depends on whether  $\Delta{\bf v}$ is non-zero for all
${\bf k}$, ${\bf p}$, and ${\bf q}$ satisfying energy conservation.
On relabeling $\mathbf{p}$ to $-\tilde{\mathbf{p}}$ and invoking time-reversal symmetry, 
the arguments of both $\delta$ functions become the same.
The problem of finding the allowed initial states ${\bf k}$ and ${\bf p}$ at fixed ${\bf q}$ now reduces to finding the solutions of the equation 
$\epsilon_{\mathbf{k}}=\epsilon_{\mathbf{k}-\mathbf{q}}$
 (and the same for $\mathbf{k}\to \tilde{\mathbf{p}}$).
Geometrically, this is equivalent to shifting the
FS by a vector $\mathbf{q}$ and finding $\mathbf{k}$ and $\tilde{\mathbf{p}}$ as the points where the original and shifted  FSs intersect.

\begin{figure}
\includegraphics[angle=0,width=0.4\textwidth]{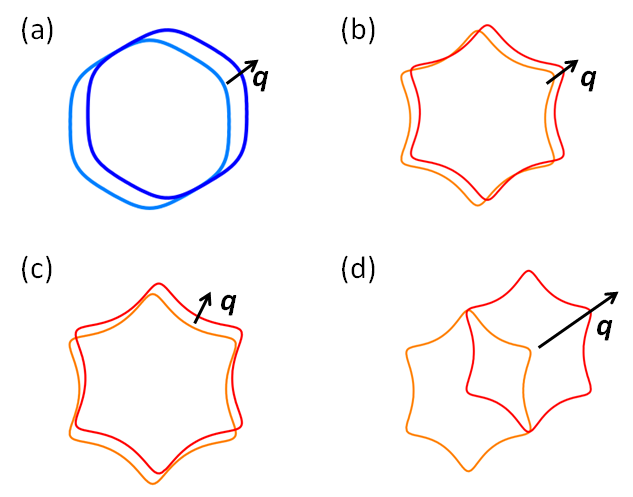}
\caption{(color online). (a) A convex contour has
no more than
two self-intersection points. (b) For $\textbf{q}$ along a 
special direction, a 
concave contour has the maximum number (6 for a sixfold-symmetric FS) of self-intersection points. 
(c)
 Even in the concave case, 
 the number of self-intersection points 
 is less than the maximal number allowed by symmetry, if $\mathbf{q}$ is not along a special direction.
 (d) For $q$ larger than a critical value, the number of the self-intersection points is less than the maximum number allowed by symmetry.}
\label{fig2}
\end{figure}

Consider first the case $\epsilon_F<\epsilon_c$, when the FS is convex. As shown in Fig. \ref{fig2}(a), there are only two points of intersection.
If $\mathbf{k}
$ is one of these intersection points then, by symmetry, the other point is $
-\mathbf{k}
+\mathbf{q}$. Since solutions for $\tilde{\mathbf{p}}$ 
must belong
to the same set $\{\mathbf{k},
-\mathbf{k}+\mathbf{q}\}$,  the scattering process either occurs  in the Cooper channel ($\{\mathbf{k},-\mathbf{k}\}\to\{\mathbf{k}-\mathbf{q},-\mathbf{k}+\mathbf{q}\}$) or corresponds to swapping of
initial momenta 
 ($\{\mathbf{k},\mathbf{k}-\mathbf{q}\}\to\{\mathbf{k}-\mathbf{q},\mathbf{k}\}$).
In both cases, $\Delta{\bf v}=0$, 
and thus the $T^2$ correction to the conductivity vanishes. The first nonvanishing term in this case is $T^4$, which can be obtained from Eq. (\ref{delsigma}) by expanding the product of the $\delta$ functions to second order in $\omega$. On the other hand, if the FS is concave ($\epsilon_F>\epsilon_c$), there are six possible points of intersection yielding six solutions for each $\mathbf{k}$ and $\tilde{\mathbf{p}}$ [cf.  Fig.~\ref{fig2}(b)].  The total set of thirty six pairs for $\{\mathbf{k},\tilde{\mathbf{p}}\}$ contains processes other than Cooper channel and swapping,
and $\Delta\mathbf{v}$ is non-zero for these processes.

The analysis presented above is valid either well below or well above the convex/concave transition, i.e., when $|\Delta|\sim \epsilon_{F}$.  We now turn to the vicinity of the transition, when $|\Delta|\ll\epsilon_F$. First, we focus on the most interesting case of $\Delta\gg T$, when the isoenergetic contours near the Fermi energy are concave,
and then discuss the case of $|\Delta|\lesssim T$, when both convex and concave contours near the FS are thermally populated.
Near the transition, several quantities in Eq.~(\ref{delsigma}) exhibit a critical dependence on $\Delta$. 
First, 
it is 
 $\Delta \mathbf{v}$ which is zero on the convex side and non-zero on the concave side. 
Additionally, 
there are two other quantities which also show a 
critical behavior. 
As Figs.~\ref{fig2}(c) and \ref{fig2}(d) illustrate, 
even if the FS is concave, it has more than two self-intersection points only if it is shifted along one of the 
special
 directions and the magnitude of the shift is sufficiently small.
 [These special directions are high-symmetry axes that intersect the FS at points with positive curvature, as in Fig. 2(b).]
  Therefore, the width of the angular interval near
a 
special direction
 ($\Delta\theta_\mathbf{q}$) and the maximum value of $q$ ($q_{\max}$ also depend on $\Delta$ in a critical manner.
 Approximating $\int d^2q$ by $\Delta\theta_\mathbf{q}q_{\max}^2$, resolving the $\delta$ functions, and integrating over 
 all
 energies, we obtain
\bea
\delta\sigma_{jj}&=&-\frac{e^2\tau_{i}^2T^2}{
12}
\sum_{l,m}
\Delta\theta_{\mathbf{q}}|M_{\mathbf{k}_{l},\mathbf{p}_{m}}(\mathbf{q}_{\max})|^2 \nonumber \\
\nonumber\\
&\times&
[\Delta \mathbf{v}_j]_{lm}^2
\frac{k_l}{\mathbf{v}_{\mathbf{k}_l}\cdot \hat{\mathbf{k}}_l}
\frac{p_m}{\mathbf{v}_{\mathbf{p}_m}\cdot\hat{\mathbf{p}}_m}\frac{1}{|\mathbf{v}^{'}_{\mathbf{k}_l}\cdot\hat{\mathbf{q}}|}\frac{1}{|\mathbf{v}^{'}_{\mathbf{p}_m}\cdot\hat{\mathbf{q}}|},
\label{delsigma3}
\eea
where 
the sum runs over all 
intersection points, the
prime denotes a derivative with respect to the azimuthal angle, and $\hat{\mathbf{l}}\equiv\mathbf{l}/|\mathbf{l}|$. 
 Notice that although a factor of $q^2_{\max}$ from the phase space of integration cancels with 
 the same factor from the $\delta$ functions, it will reappear in the calculation of $\Delta\mathbf{v}_j$.
 We are now going to show that
  \begin{equation}
 \Delta\theta_{\mathbf{q}}\propto \Delta^{3/2}, q_{\max} \propto \Delta^{1/2},\; \mathrm{and}\; \Delta\mathbf{v}_j\propto\Delta^{3/2}.
\label{eq:scaling} \end{equation}

\begin{figure}
\includegraphics[angle=0,width=0.5\textwidth]{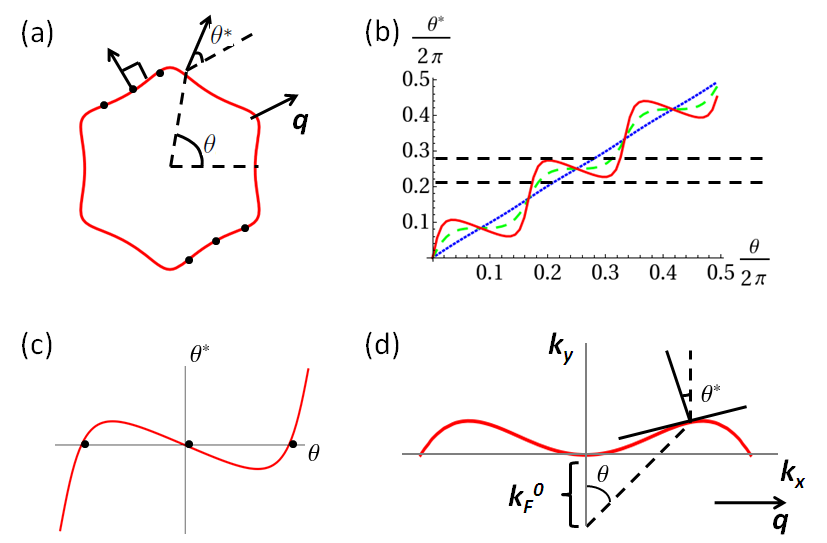}
\caption{(color online). (a) For small $q$, points where the normal to the FS is perpendicular to $\mathbf{q}$ are the points of self-intersection (black dots). 
(b) $\theta^*$ vs $\theta$ [as defined in panel (a)]. Dotted: $\epsilon<\epsilon_c$; dashed: $\epsilon=\epsilon_c$; solid: $\epsilon>\epsilon_c$. 
(c) A zoom of the non-monotonic part of the graph in panel (b).  
(d) A portion of the FS contour showing the geometrical construction for the derivation of the equation for the contour.}
\label{fig3}
\end{figure}

We begin with $\Delta\theta_{\mathbf{q}}$. Under an assumption (to be justified later) of small $q$, the equation $\epsilon_{\mathbf{k}}-\epsilon_{\mathbf{k}-\mathbf{q}}=0$ reduces to $\mathbf{v}_{\mathbf{k}}\cdot\mathbf{q}=0$, which implies that $\mathbf{q}$ is a tangent to the FS at the intersection points [cf. Fig.~\ref{fig3}(a)]. 
 Defining $\theta^*$ as an angle between the normal to the FS at any given point and 
 $\mathbf{q}$, we 
  plot $\theta^*$as a function of the azimuthal angle $\theta$.
 Figure \ref{fig3}(b) clearly demonstrates a distinguishing feature between the 
convex and concave contours: $\theta^*(\theta)$ is monotonic for the former and non-monotonic for the latter.
 The non-monotonic part 
 is centered around 
certain invariant points, i.e., common points 
for all contours.
The oscillations
reflect the rotational symmetry--sixfold in our case--of the FS.
From symmetry, if $\theta$  is a solution, so is $\theta+\pi$; we thus consider only the domain $\theta\in [0,\pi]$. 
We now need to find the angular interval of $\mathbf{q}$
about a special direction
 in which
the equation $\theta^*=\theta_{\mathbf{q}}+\pi/2$ has three roots;
the $T^2$ term is non-zero only in this case.
Since 
there is a one-to-one correspondence between the angles $\theta^*$ and  $\theta_{\mathbf{q}}$, we can find the corresponding interval 
$\Delta\theta^*$ instead of  $\Delta\theta_{\mathbf{q}}$. Clearly, the regions on the curve where it is non-monotonic 
are
responsible for the multiple roots \cite{light}.
Redefining variables $\theta$ and $\theta^*$ as 
measured from the invariant points, the non-monotonic part of the curve can be conjectured
to obey a cubic equation [cf. Fig. \ref{fig3}(c)]:
\beq
\theta^*=b\theta^3-a(\Delta)\theta,
\label{cubic}
\eeq
where $a(\Delta)\propto \Delta
$ and $b>0$ is a constant.
Indeed, we need at least a cubic equation to provide for three real roots; whether there is one or three roots depends on the sign of $a(\Delta)$ which must be negative/positive in the convex/concave regimes, correspondingly.
For the model spectrum of Eq.~(\ref{eq:spectrum}), we find $b=2$ and $a(\Delta)=\frac{16}{9}\sqrt{\frac{7}{6^{1/2}}(\frac{\lambda}{v^3})}\Delta$.
The quantity 
 $\Delta\theta^*$ is the 
 vertical distance 
 between the maximum and minimum of this curve which,
 according to Eq.~(\ref{cubic})
 scales as $\Delta
 ^{3/2}$.
 
 Next in line is $\Delta\mathbf{v}_j$,
 which we expand in small $q$ as
 $\Delta\mathbf{v}_j\approx [\frac{\partial}{\partial k_j}(\mathbf{q}\cdot\mathbf{v})]|_{\mathbf{k}_2}-[\frac{\partial}{\partial k_j}(\mathbf{q}\cdot\mathbf{v})]|_{\mathbf{k}_1}=[\frac{\partial}{\partial k_j}(\mathbf{q}\cdot\mathbf{v})]|^{\mathbf{k}_2}_{\mathbf{k}_1}$, where $\mathbf{k}_2$ and $\mathbf{k}_1$ are any two solutions of the equation $\mathbf{q}\cdot\mathbf{v}_\mathbf{k}=0$. Referring to the geometry of Fig. \ref{fig3}(d), we find
 $\Delta\mathbf{v}_j\propto q[\frac{\partial}{\partial\theta}(\theta^*)]|^{\theta_2}_{\theta_1}$, where use of Eq.~(\ref{cubic}) yields $\Delta\mathbf{v}_j\propto q
 \Delta\approx q_{\max}\Delta
 $. 
 
 Finally, to find  $q_{\max}$, 
 we relax the assumption
 of small $q$
  and solve the 
  equation $\epsilon_{\mathbf{k}-\mathbf{q}}=\epsilon_{\mathbf{q}}$ for arbitrary $q$. 
  It is easier to do this by casting Eq.~(\ref{cubic}) 
  into an equation for the contour in terms of local cartesian coordinates. To this effect, we approximate $\theta^*\approx \mathrm{tan}\theta^*\approx dk_y/dk_x$ and $\theta \approx \mathrm{tan}\theta\approx -k_x/k_F^0$, with $k_F^0$ being the Fermi momentum at the invariant point [cf. Fig.~\ref{fig3}(d)] and substitute into Eq.~(\ref{cubic}) to get the following contour equation: 
$k_y=-b k_x^4/4+a(\Delta)k_x^2/2$, where 
   $k_{x,y}$ are the momenta 
  measured from the invariant points 
  and  normalized 
  by $k_F^0$.
  Using this expression to solve for 
  the roots of
  $\epsilon_{\mathbf{k}-\mathbf{q}}=\epsilon_{\mathbf{q}}$ 
  one arrives at a cubic equation in $k_x$,
  which has three distinct real roots
  if $q\le 2\sqrt{a(\Delta)/b}$. This means that 
  $q_{\max}\propto 
  \Delta^{1/2}
  $ (and thus $\Delta {\bf v}_j\propto \Delta^{3/2}$). This, in hindsight, validates the assumption of small
  $q$.

Collecting all the terms together, the resultant energy dependence of the critical terms is $
\Delta\theta_{\mathbf{q}}[\Delta\mathbf{v}_j]^2
\propto
\Delta
^{9/2}$. 
which leads to the scaling form of the prefactor of the $T^2$ term in Eq.~(\ref{eq:result}).
The Cooper and swapping channels of scattering always contribute a $T^4$ term to the resistivity,
and the second term in Eq.~(\ref{eq:result}) accounts for this contribution.
A crossover between the $T^4$ and $T^2$ behaviors occurs at $T\sim\epsilon_F(\Delta/\epsilon_F)^{9/4}\ll\epsilon_F$.  
A large value of the exponent ($9/2$) indicates that one needs to go sufficiently 
high
above the convex/concave transition in order to see the $T^2$ term.

Now we return to the range of energies $\Delta\lesssim T$,
when both convex and concave contours are populated.
Instead of a $\Delta^{9/2}$ factor, we get 
a factor of $T^{9/2}$,
leading to a $T^{13/2}$ term in $\rho$. This term, however, is subleading to the $T^4$ one.
Therefore, we conclude that Eq.~(\ref{eq:result}) describes the leading $T$-dependence of the resistivity in all possible situations near the transition.

Next, we discuss the feasibility of observing these predictions in an experiment. 
First of all, 
one needs to ask 
if the {\em e-ph} contribution to the resistivity masks
the {\em e-e} one.
 In general,  the {\em e-ph} contribution, which scales as $T^5$ at $T<T_{BG}$, where $T_{BG}$  is Bloch-Gruneisen temperature ($\approx 10 K$ for Bi$_2$Te$_3$ \cite{eph}),
 is expected to be outweighed by the $T^2$ (or even $T^4$) one from the {\em e-e} interaction.
 However, the {\em e-e} coupling 
 may 
 be 
 substantially 
  reduced
 due to high background polarizability of TI materials.
Indeed, comparing the scattering time of {\em e-ph} interaction, calculated in Ref.~\cite{eph},  with that of the {\em e-e} interaction, we find that the $T^5$ term
 dominates over the $T^2$
 one
 down 
 to a few mK. 
 Even with some uncertainty in the estimate of the {\em e-ph} time related to screening of this interaction by free electrons, 
the detection of the $T^2$ term in a {\em dc} measurement seems to be difficult.
 Instead, as an alternate route to test our predictions, we suggest measuring the optical conductivity as a function of the frequency $\Omega$. Indeed, the crossover between the $T^4$ and $T^2$ forms of the {\em dc} resistivity is completely analogous to the that between the $\Omega^4$ and $\Omega^2$ forms of the optical scattering rate $\Gamma(\Omega)\equiv(\omega^*_p)^2\mathrm{Re}\rho(\Omega)$ \cite{rosch}.
It can be readily shown that the effective plasma frequency $\omega_p^*$ contains the same integrals  as in Eq. (\ref{delsigma})  \cite{rosch,plasma}.
Therefore, all the foregoing conclusions on the temperature dependence at different values of $\epsilon_F$ carry over to the frequency dependence. The advantage of an optical measurement is that
the {\em e-ph} part of $\Gamma(\Omega)$ saturates \cite{ephsaturate}
for $T_{BG}\ll \Omega\ll\epsilon_F$,
while  
the {\em e-e}  part 
continues to grow
either as $\Omega^2$ or $\Omega^4$, depending on the sign of $\Delta$. This advantage was used in the past to detect the {\em e-e} contribution to $\Gamma(\Omega)$ in noble metals \cite{ephsaturate}, and we propose to apply the same technique to TIs.

\acknowledgments
This work was supported by NSF-DMR-0908029 (H.K.P. and D.L.M. ) and RFBR-09-02-1235 (V.Y.). 
D.L.M. acknowledges the hospitality  of MPI-PKS (Dresden), where a part of this work was done.


\begin{thebibliography}{99}
\bibitem{comment2} Although we use a scalar BE instead of a spinor one suitable for helical states, it does not affect any of our conclusions. 
Away from the Dirac point, the effect of the phase factors in the spinor wavefunctions is
 only to suppress backscattering probability of all processes considered.
\bibitem{light}In light of this, it is now clear why a convex contour does not allow for more than two solutions and  why a concave 
contour allows for more than two solutions only for special directions of $\mathbf{q}$.
\end{thebibliography}
\end{document}